\documentclass[a4paper,twoside]{article}

\usepackage{epsfig}
\usepackage{subcaption}
\usepackage{calc}
\usepackage{amssymb}
\usepackage{amstext}
\usepackage{amsmath}
\usepackage{amsthm}
\usepackage{multicol}
\usepackage{pslatex}
\usepackage{apalike}
\usepackage{algorithm2e}
\usepackage[bottom]{footmisc}
\usepackage[disable]{todonotes}
\usepackage{hyperref}
\usepackage{url}
\usepackage{booktabs}
\usepackage{svg}
\usepackage{SCITEPRESS}     

\usetikzlibrary{tikzmark}

\newcommand{\overlaycircle}[2]{%
	\tikz[overlay,baseline=(char.base)]\node[anchor=north, draw=#1, circle, inner
	sep=1pt,fill=#1,text=white, scale=0.9, minimum size=13pt, inner sep=0pt,font=\sffamily,
	solid](char){#2} ;}
\newcommand{\circled}[2]{%
	\tikz[baseline=(char.base)]\node[anchor=north, draw=#1,circle, inner
	sep=1pt,fill=#1,text=white, scale=0.9, minimum size=13pt, inner sep=0pt,
	solid](char){#2} ;}

\begin{document}

\title{EMERALD: Evidence Management for Continuous Certification as a Service in the Cloud}

\author{\authorname{Christian Banse\sup{1}\orcidAuthor{0000-0000-0000-0000}, Björn Fanta\sup{2}\orcidAuthor{0000-0000-0000-0000}, Juncal Alonso\sup{3}\orcidAuthor{0000-0002-9244-2652} and Cristina Martinez\sup{3}\orcidAuthor{0000-0002-3302-1129}}
\affiliation{\sup{1}Fraunhofer AISEC , Garching b. München, Germany}
\affiliation{\sup{2}Fabasoft, Linz, Österreich }
\affiliation{\sup{3}TECNALIA, Basque Research and Technology Alliance (BRTA), Derio, Spain}
\email{banse@aisec.fraunhofer.de, bjoern.fanta@fabasoft.com,juncal.alonso@tecnalia.com, cristina.martinez@tecnalia.com}
}


\abstract{The conspicuous lack of cloud-specific security certifications, in addition to the existing market fragmentation, hinder transparency and accountability in the provision and usage of European cloud services. Both issues ultimately reflect on the level of customers' trustworthiness and adoption of cloud services. The upcoming demand for continuous certification has not yet been definitively addressed and it remains unclear how the level 'high' of the European Cybersecurity Certification Scheme for Cloud Services (EUCS) shall be technologically achieved. The introduction of AI in cloud services is raising the complexity of certification even further. This paper presents the EMERALD Certification-as-a-Service (CaaS) concept for continuous certification of harmonized cybersecurity schemes, like the EUCS. EMERALD CaaS aims to provide agile and lean re-certification to consumers that adhere to a defined level of security and trust in a uniform way across heterogeneous environments consisting of combinations of different resources (Cloud, Edge, IoT). Initial findings suggest that EMERALD will significantly contribute to continuous certification, boosting providers and users of cloud services to maintain regulatory compliance towards the latest and upcoming security schemes.}

\onecolumn \maketitle \normalsize \setcounter{footnote}{0} \vfill

\section{\uppercase{Introduction}}
\label{sec:introduction}
Cloud-based services have grown from basic computing services to complex ecosystems, comprising (virtual) infrastructure, business processes, and application code. These advanced services also increasingly leverage the usage of Artificial Intelligence, including Machine Learning or Natural Language Processing techniques, raising the complexity even higher. Due to the cascade of dependencies between different products and services, the need has arisen to make the certification process for cloud-based services more agile, for example by using continuous monitoring and assessment, as evidenced by references to it in the EU Cybersecurity Act \cite{EUCSA} (EU CSA) certifications. 

To transform the continuous assessment and certification concept into the complete realization of a Certification-as-a-Service (CaaS), several challenges need to be solved: 1) the state-of-the-art proofs of concept for continuous monitoring lack interoperability at technological level\todo{CB: I dont know what you mean with this JA: Lack of interoperability between tools to gather different types of evidences, process them, etc.}, 2) the adoption of cloud and edge computing and the incorporation of topic- or domain-specific regulations, such as AI, involves a significant strain on companies to comply with a multitude of different security schemes, 3) the existing market fragmentation for continuous certification hinders transparency and accountability in the provision of European cloud services, and 4) smart tools and models need to be adopted to ease the agile application and implementation of the CaaS concept, reducing complexity in the whole cloud certification value chain and facilitating the adoption of CaaS by the various stakeholders.

In this paper, we present \textbf{EMERALD}, a novel approach to automatic cloud service certification with a focus on evidence management. The main objective of EMERALD is to provide a framework that enables continuous Certification-as-a-Service (CaaS) and agile and lean re-certification. Targets are services that adhere to a defined level of security and trust in a uniform way across heterogeneous environments made of combinations of various Cloud and IoT resources. 

\section{\uppercase{Continuous Cybersecurity Certification}}
\label{sec:section2}
Continuous cybersecurity certification is a concept inspired by the \textit{“Continuous Auditing”} notion mentioned by ENISA during the creation of the EUCS \cite{eucs}. It refers to cybersecurity requirements related to continuous monitoring, with the intended meaning of \textit {“automatic monitoring”}. This involves 1) gathering data at discrete intervals with sufficient frequency, 2) 	comparing the gathered data to a reference, 3) reporting deviations for timely analysis, 4) initiating a process to fix any non-conformity discovered and 5) notifying the CAB (Conformity Assessment Body) of a major non-conformity. 

Continuous certification offers significant advantages by providing an ongoing evaluation and auditing process, unlike the current certification process, which is typically conducted in larger fixed interval, e.g., one year. In the current process, cybersecurity requirements are assessed and audited in a discrete manner. In contrast, continuous certification allows internal or external auditors to perform certification-related activities on a more continual basis.

Achieving continuous cybersecurity certification requires overcoming significant challenges in interoperability, regulatory coherence, and evidence reuse. Frameworks like EMERALD, which build upon the findings of projects such as MEDINA \cite{medina2021}, offer promising solutions by introducing automated evidence collection, certification graphs, and adaptive compliance mechanisms.
The EU has acknowledged these challenges through initiatives like the EU CSA, promoting continuous certification methodologies. However, despite technological advancements, European companies often face barriers to entry, whether as consumers or providers of cloud services. Lack of interoperability, market fragmentation, and the absence of comprehensive, reusable evidence frameworks are significant hurdles that limit trust and thus participation and growth in the cloud ecosystem.

\subsection{Context and Need}
Cloud computing services have become indispensable across industries, with advanced functionalities such as machine learning (ML) and natural language processing (NLP) being integral to modern applications. According to Eurostat, the adoption of cloud services in large enterprises increased by 21 percentage points since 2014, highlighting a paradigm shift in operational frameworks. Cloud-based systems now encompass intricate layers of infrastructure, business processes, and application code, amplifying the need for robust security measures.

\subsection{Challenges and Gaps}
The transition to continuous cybersecurity certification is fraught with several challenges, ranging from technological interoperability to regulatory fragmentation. In the following, we list four key issues.

\begin{enumerate}
    \item Fragmentation in Certification Schemes: The co-existence of various certification schemes, such as ISO standards, the Cloud Security Alliance’s Cloud Controls Matrix (CCM) \cite{ccm}, and country-specific frameworks (e.g., German BSI C5, Spanish ENS, French SecNum Cloud), complicates compliance efforts. The EU Cybersecurity Certification Scheme for Cloud Services (EUCS) by \cite{eucs}, aims to address this fragmentation but lacks detailed implementation guidelines for achieving high-assurance levels.
    \item Interoperability Challenges: Cloud systems rely on a diverse range of tools and technologies, creating interoperability issues in continuous monitoring and assessment. The Open Security Controls Assessment Language (OSCAL) originally developed by \cite{piez2019open} offers potential solutions but has not gained widespread adoption (especially in Europe), leading to inconsistencies in data formats and evaluation methods.
    \item Stakeholder Specific Barriers: Both, consumers and providers of cloud services, face disproportionately high entry barriers. As consumers, stakeholders struggle with limited expertise and resources to secure operations effectively. As providers, they often lack visibility and face challenges in integrating with larger systems, exacerbating interoperability and compliance difficulties.
    \item AI Integration Complexities: The integration of AI technologies, such as LLM, ML and NLP, further complicates certification processes. AI models require specialized evaluation methods to ensure robustness against adversarial attacks, bias, and data poisoning. While frameworks like the AI Cloud Service Compliance Criteria Catalogue (AIC4) address these concerns, they remain nascent and fragmented.
\end{enumerate}

Current and future research must focus on harmonizing certification, establishing standards for evidence management, fostering stakeholder inclusiveness, and addressing AI-specific challenges.

\section{\uppercase{Towards Certification-as-a-Service (CaaS): The EMERALD approach}}
\label{sec:approach}

\begin{figure}
    \centering
        \begin{tikzpicture}[      
        every node/.style={anchor=north west},
        x=1mm, y=1mm,
      ]   
     \node (fig1) at (0,0) {\includegraphics[width=0.9\linewidth]{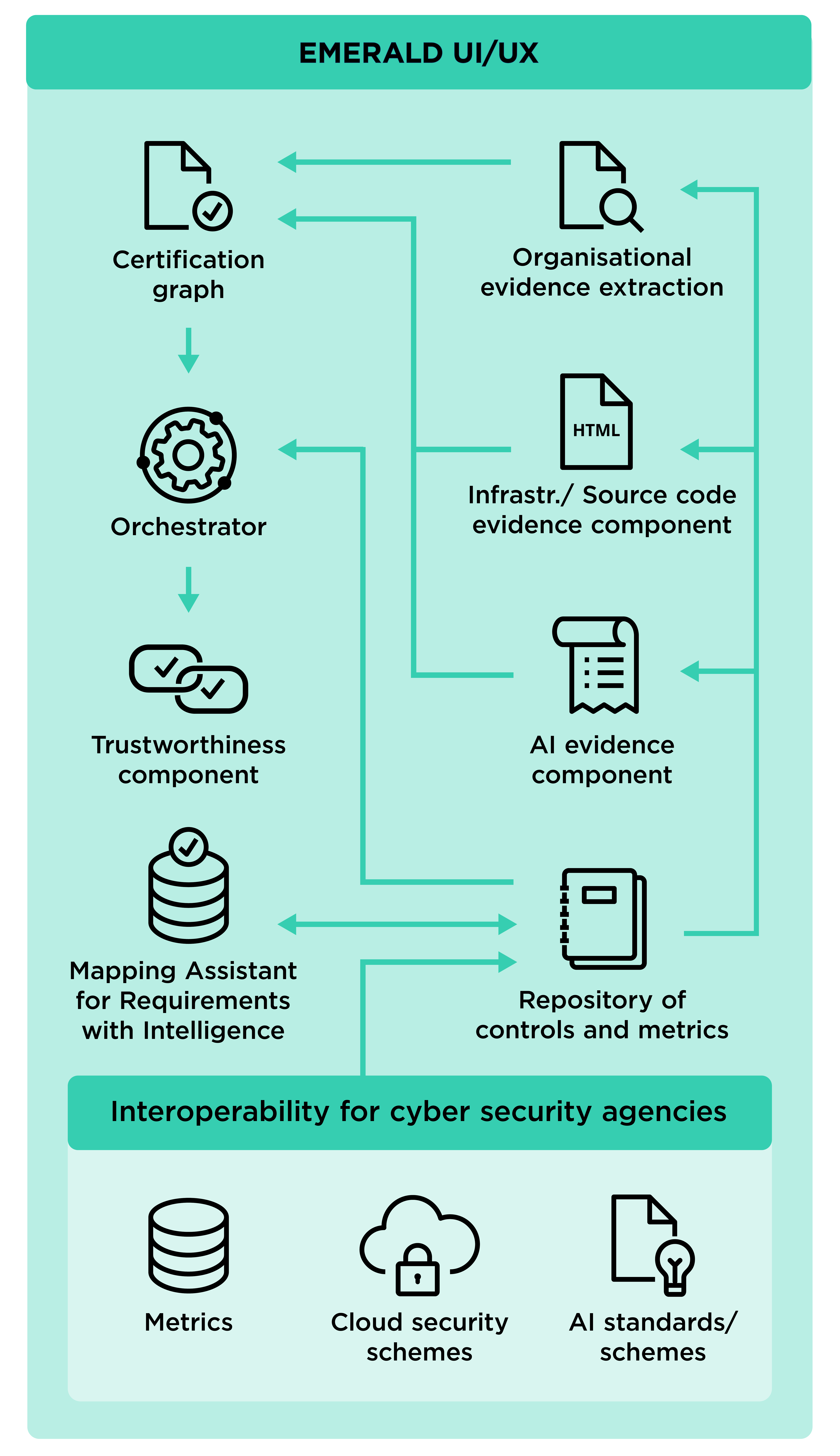}};
     \node at (57,-68) {\overlaycircle{black}{1b}};
     \node at (9,-68) {\overlaycircle{black}{1c}};
     \node at (26,-94) {\overlaycircle{black}{1a}};
     \node at (57,-8) {\overlaycircle{black}{2c}};
     \node at (57,-28) {\overlaycircle{black}{2b}};
     \node at (57,-48) {\overlaycircle{black}{2a}};
     \node at (19,-8) {\overlaycircle{black}{3}};
     \node at (19,-28) {\overlaycircle{black}{4a}};
     \node at (19,-48) {\overlaycircle{black}{4b}};
    \end{tikzpicture}
    \caption{The EMERALD approach showing definition of  controls and metrics as well as the collection and assessment of evidence in the certification graph. 
    }
    \label{fig:architecture}    
\end{figure}
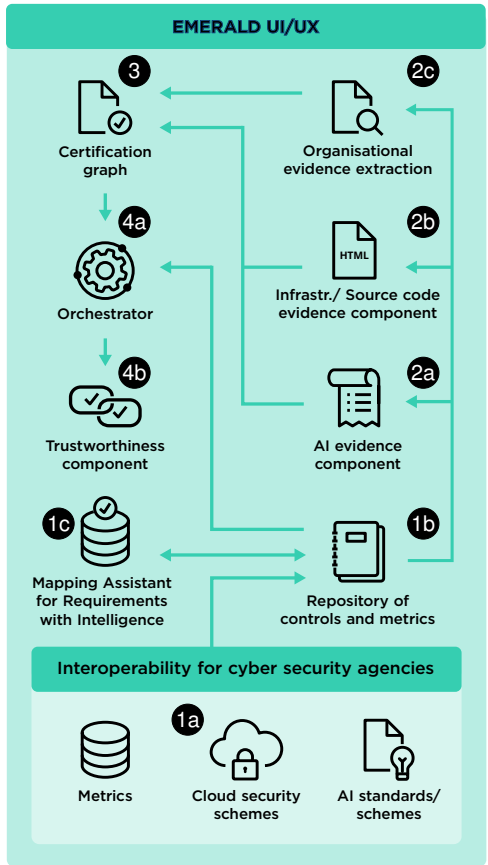

The EMERALD approach, as seen in Figure~\ref{fig:architecture} is driven by the typical workflow encountered when dealing with certification and auditing, and ranges from extracting evidence and storing evidence, managing the necessary meta-information about the targeted controls to the evaluation and assessment of the selected controls according to a selected schema. In the following, the individual parts will be detailed.

\subsection{Representing Standards, Controls and Metrics}
\label{sec:rcm}

\paragraph{Repository of Controls and Metrics} In order to automatically demonstrate compliance to security catalogs and certifications, these works need to be available in a machine-readable form. A \textit{repository of controls and metrics} (\circled{black}{1b}) is the central component to hold these information \cite{rcm}.
EMERALD promotes interoperability with standards such as OSCAL in order to provide data such as security schemes and certifications as well as the controls defined by them (\circled{black}{1a}). Furthermore, EMERALD proposes scheme-independent security metrics in order to provide a generic security assessment of the target system (called \textit{certification target}), which can then be put into the context of a particular audit or certification. This is known as an \textit{audit scope} within the framework. Table~\ref{tab:metric} shows an example of the metadata defined for each metric.

\begin{table}
    \centering
    \caption{Example metadata of metric \textit{TransportEncryptionProtocolVersion}.}
    \begin{tabular}{p{2cm}p{3.5cm}}
        \textbf{Key} & \textbf{Value} \\
        \toprule
         Name & Transport Encryption Protocol Version \\
         \midrule
         Description & This metric is used to assess if an up-to-date transport encryption protocol version is used. \\
         \midrule
         Category & Transport Encryption \\
         \midrule
         Scale / Values & Ordinal, [1.0, 1.1, 1.2, 1.3] \\
         \midrule
         Recommended Target Value & $>= 1.2$ \\
         \midrule
         Interval & on-demand, every 5 minutes \\
         \bottomrule
    \end{tabular}
    \label{tab:metric}
\end{table}

\paragraph{Metric Recommendation and Mapping}

The mapping between controls of a particular catalog to one or more metrics is not static. Over the course of the development, it is expected that new metrics arise that may be more suitable than existing ones. Furthermore, security schemes are also frequently updated. Therefore, we envision a component that selects the most suitable set of metrics for each control in the catalog in an intelligent way (\circled{black}{1b}).
Since the re-usability of evidence (and therefore metrics) is one of the key goals for EMERALD, different strategies of choosing metrics can be imagined. One of the ideas is to combine security schemes of different granularity and domains. For example, a company might choose to target the BSI C5 to ensure its baseline cloud security. But since it is also employing AI in its cloud service, it aims to also be compliant to an upcoming AI security scheme. Since there is a potential overlap in both schemes when it comes to ensuring the security of data, a set of common metrics that are suitable for both schemes should be chosen.

The recommendation and mapping of metrics is explored by the EMERALD component \textit{MARI (Mapping Assistant for Requirements with Intelligence)} (\circled{black}{1c}), which takes text inputs (e.g., strings) and analyses, whether the strings are \textit{similar} or \textit{equal enough}. 
MARI enables a user to time-efficiently map different cybersecurity schemes among each other. (1) a user is able to find fitting (already implemented) metrics for controls. (2) a user is able to map controls of different new target schemes to controls of a selected source scheme. In case of (1), the \textit{association of controls and metrics} takes as an input one particular control and a set of metrics. The output is a list of metrics relevant for the control, ordered according to relevancy. For the case (2), \textit{association of controls among different schemes}, the input is one control of one scheme and a set of controls of another scheme. The result is a list of a list of controls (of the target scheme), ordered according to relevancy to the source control.



\subsection{Extraction and Modeling of Evidence}

At the core of the EMERALD framework, the concept of \textit{evidence} is used. An evidence is a piece of information that proves that a system poses certain properties or behaves in a certain desired way \cite{anisetti2020}. EMERALD builds on the concept of so-called \textit{semantic evidence} as defined by \cite{banse2023semanticevidence}. This extends the original evidence definition by \cite{anisetti2020} by a structure defined in an ontology or taxonomy. We aim to extend the mentioned previous work on integrating semantic technologies into the cloud-certification process and provide an extensive ontology of all concepts related to the certification process. The objective is to unify all this information into a common knowledge graph, the \textit{Certification Graph} (\circled{black}{3}) \cite{schoeberl2024}. EMERALD envisions the collection, using so-called \textit{evidence extractors} (\circled{black}{2a}, \circled{black}{2b}, \circled{black}{2c}), and semantic modeling of such evidence at different layers.

\paragraph{Infrastructure Layer} The (virtual) infrastructure layer includes different types of resources deployed in the cloud, their properties as well as their relationships. Evidence of configurations will be extracted using open-source tools such as Clouditor\footnote{\url{https://github.com/clouditor/clouditor}}, as well as by providing interfaces to commercial cloud security posture tools or native solutions such as Azure Policy\footnote{\url{https://learn.microsoft.com/en-us/azure/governance/policy/overview}}.

\paragraph{Application Layer} Cloud services not only comprise the infrastructure layer, but are usually also made of applications and business code. It is of utmost importance that not only the resources deployed in the cloud, but also the code managing the service itself adheres to the principles of security standards and certifications. The EMERALD approach also considers this by including a rich semantic description of the applications deployed on or interacting with infrastructure resources. This also comprises a classification of the functional or behavioral patterns of the application, for example whether the application interacts with a database or issues HTTP(S) requests. Code property graphs \cite{yamaguchi2014cpg,weiss2022languageindependentanalysisplatformsource} or commercial tools 
can be leveraged to collect evidence on this layer. 

\paragraph{Organizational Layer} Often, security standards and certifications refer not only to technical measures but also to policies and procedures that must be in place. Therefore, it is important to also consider the organizational layer, usually comprised of documents describing said procedures. The proposed certification graph should include a semantic understanding of the different types of documents; automated assessment using techniques such as Large Language Models (LLMs) and Natural Language Processing (NLP) is necessary in order to extract compatible evidence out of the documents. One of the major challenges is that EMERALD aims to harmonize evidence gathered from documents with evidence gathered from technical layers by using a common set of metrics for both. Instead, previous approaches such as \cite{deimling2023amoe} relied on a separate set of metrics for analyzing documents.

\paragraph{Data Layer} The data layer describes the actual business data processed by the cloud service. With the recent advancements in AI, a secure storage and processing of AI models, such as LLMs becomes paramount. Therefore, we aim to include a classification of AI models and parameters in order to make a statement about different metrics of an AI model, such as fairness or robustness.

\begin{figure*}
    \centering
    \scalebox{0.38}{
    \input{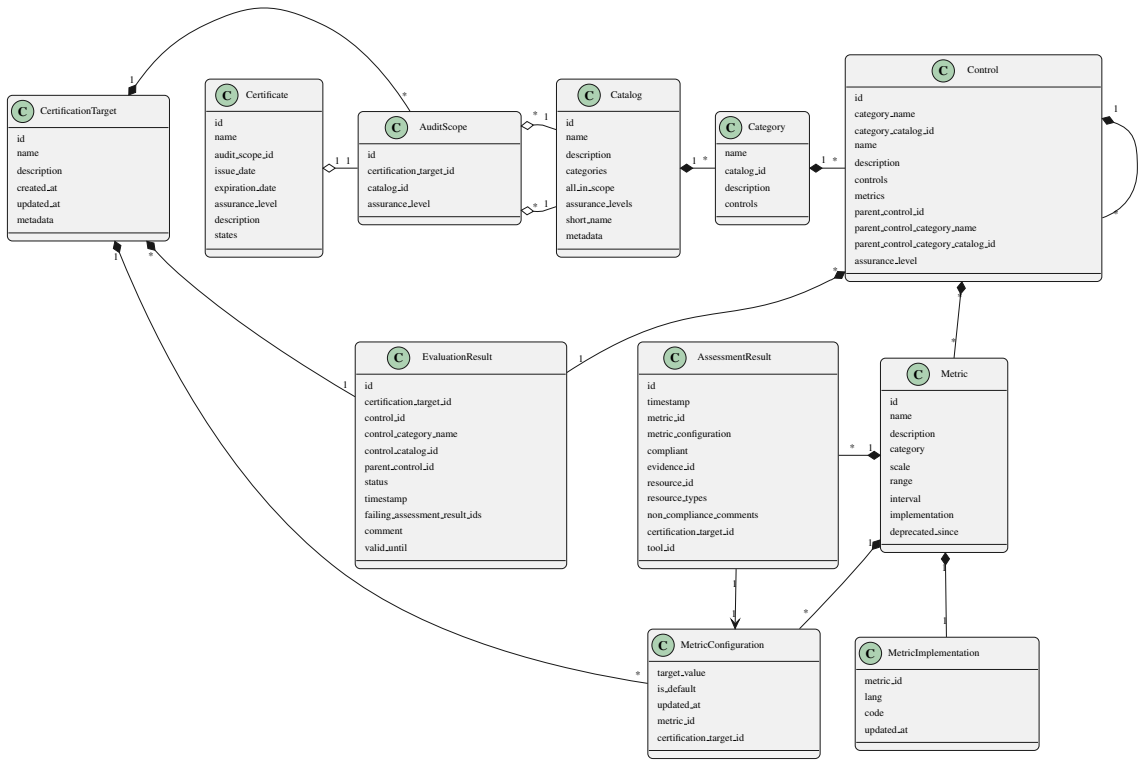}
    }
    \vspace{1em}
    \caption{Excerpt of the data model used by various EMERALD components.}
    \label{fig:data-model}
\end{figure*}

\subsection{Orchestration, Assessment and Trustworthiness}

The \textit{orchestrator} (\circled{black}{4a}) component is in charge of overseeing the whole certification process. It keeps the state of the various evidence collectors and schedules the necessary evaluation and assessment of evidence in the certification graph, to finally arrive at a certification decision. 

Figure~\ref{fig:data-model} shows an excerpt of the common data model used by the orchestrator and the remaining components.
\begin{itemize}
    \item The \textit{Catalog}, \textit{Category} and \textit{Control} classes are used to model data related to the security catalogs and schemes (see Section~\ref{sec:rcm})
    \item The \textit{Metric} class represents a security metric and can be further described by its \textit{MetricImplementation}. The implementation of a metric can be done in different programming languages. In EMERALD we leverage the logic programming language Rego\footnote{\url{https://www.openpolicyagent.org/docs/latest/policy-language/}}.
    \item The \textit{CertificationTarget} represents the target or system we want to assess. In the EMERALD context, this is usually a cloud service, but with the advent of regulations like NIS-2 or the Cyber Resilience Act (CRA), we aimed to chose a neutral name for this class. Each service can configure specific target values for metrics (in the form of a \textit{MetricConfiguration}), to account for company-specific security defaults.
    \item As part of the assessment, an \textit{AssessmentResult} is created for each metric, based on suitable evidence by querying the certification graph.
    \item In order to put the results of the assessment in the context of a concrete audit, an \textit{AuditScope} is created. This comprises a certification target and a selected catalog. For each control of the selected catalog, an \textit{EvaluationTarget} is created by combining the results of suitable assessment results -- based on the mapping of metric to control. 
    \item Finally, the state of a \textit{Certificate} object can be updated based on the evaluation results: either the certificate is still valid if all assessment results are ``ok'', otherwise a minor or major deviation is detected and the certificate owner must take actions to restore it to a healthy state.
\end{itemize}

\paragraph{Ensuring Trust in the Approach}

\begin{figure*}
    \centering
    \includegraphics[width=\linewidth,trim={0 5cm 0 0},clip]{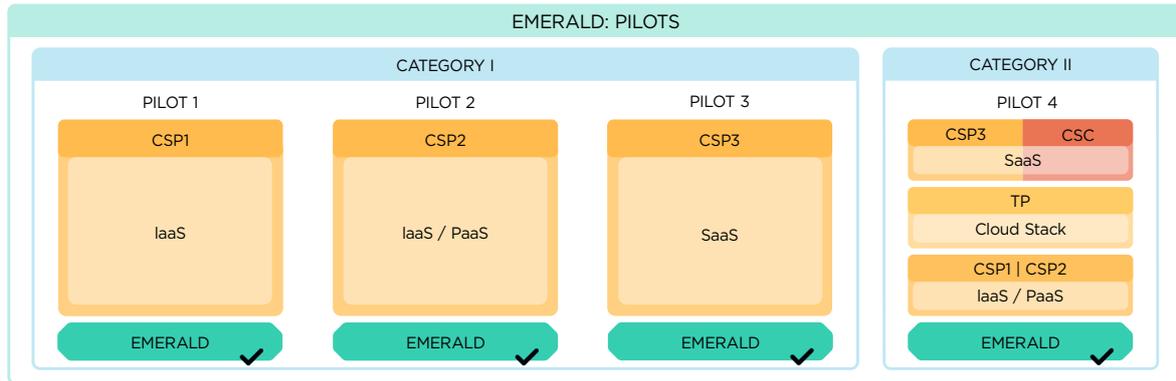}
    \caption{Overview of the four EMERALD pilots, involving three cloud service providers (CSP1-3), one cloud service customer (CSC) as well as one technology provider (TP). The individual pilots are focused on different cloud service models, e.g., IaaS/PaaS/SaaS.}
    \label{fig:pilots}
\end{figure*}

The \textit{trustworthiness component} (\circled{black}{4b}) is in charge of ensuring the integrity of all evidence results through the use of Blockchain technology as backbone, and can be queried to find out whether data in the process was manipulated. To ensure the integrity of the entire evidence collection and processing chain, from the initial \textit{evidence extractors} (\circled{black}{2a}, \circled{black}{2b}, \circled{black}{2c}) to the final assessment by the \textit{orchestrator} (\circled{black}{4a}), we are exploring the creation of a hash for each piece of evidence that avoids evidence disclosure, so that the hashes will be transmitted signed from the evidence sources to the \textit{trustworthiness component} (\circled{black}{4b}) for recording. This feature will be optional, allowing each \textit{evidence extractor tool} to choose whether to record the hashes directly in the \textit{trustworthiness component }(\circled{black}{4b}) or via the \textit{evidence store} in the orchestrator (\circled{black}{4a}). While the second option is already implemented, we are currently assessing the implementation of this approach through a proof of concept for the first option. The technical design and implications of this method are under discussion.

\section{\uppercase{Towards Validation}}
\label{sec:validation}

In the following we give a brief overview on the current state of implementation and the validation approach used in EMERALD. We use different pilots to validate the scientific and technological results of the project and to address the usability aspect of the EMERALD approach towards CaaS.

\subsection{Current State of Implementation}

We are currently in the process of implementing the individual components of the EMERALD framework as an open-source solution\footnote{\url{https://git.code.tecnalia.com/emerald/public}}. Because of potential different technical requirements, the technology stack of each component includes Java, Python and Go. The components work together as micro-services, which communicate either using REST or gRPC, depending on the amount of data exchanged. In each case, well defined APIs using OpenAPI or protobuf definitions ensure a smooth data exchange between all components.

The current state of implementation at the time of writing comprises a first preliminary version of all components, with the aim of deploying the components in an integrated development cluster. Once the integration is finished, we aim to validate various aspects of our approach using several pilots with different partners.

\subsection{Real-world Pilots in EMERALD}

In order to validate the proposed framework, EMERALD features four pilots in two different categories (see Figure~\ref{fig:pilots}). The challenges identified in Section~\ref{sec:section2} are addressed in these two categories. In fact, the challenge for a cloud service provider in these domains is to be able to provide services not only to adhere to the mentioned security and privacy standards, but also to specific regulatory policies and business policies specific for the sector, and also to ensure the end users a correct, coherent and transparent processing of information and data over the whole certificate life-cycle.
The four pilots are mainly characterized by closed cloud-ecosystems, this implies that in the world of enterprise applications, they have a high entry barrier for SMEs and small service providers - unless they specialize their service individually for a particular customer, which does not scale.

\paragraph{Category I - Certification of Private Clouds} The first three pilots aim for demonstrating CaaS on the level of cloud services (IaaS, PaaS and SaaS). Pilots of Category I set their focus to public cloud environments and will build upon the findings and results already achieved in MEDINA\footnote{\url{https://medina-project.eu}}. Pilots of Category I will target compliance to the level ‘high’ for continuous  certification with the EUCS and make use of a developed EMERALD UI, a component that is currently under develoment.

\paragraph{Category II – Certification of Hybrid Multi-Clouds} The fourth pilot aims to certify hybrid cloud-edge environments for the financial sector. Due to regulation, there is a pressing need for continuous certification in this sector. The application of EMERALD would ensure real-time assessment of several cloud services, validating that they are compliant with the controls defined in a specific security framework. Category II pilot will also target compliance to the level ‘high’ for continuous certification with the EUCS. The specificity of Category II is that the EMERALD approach can provide a platform to exchange real-time information on the certification states for services within the datacenter-cloud-edge continuum used in the financial sector. More specifically, it offers a secure-by-design application that monitors compliance of services with the same technology on-prem, on the cloud, or at the edge (public or private). This ensures the secure integration of third-party services, guaranteeing their validation of fit-for-purposes in the light of the Digital Operational Resilience Act (DORA).

\paragraph{Expected Benefits from the Pilots} The main goal of the pilots in both categories is to validate the concepts for a CaaS framework approach of EMERALD and propose direct insights and valuable feedback along the technical implementation phase of EMERALD.

\section{\uppercase{Related Work}}
\label{sec:related}

This section provides a short overview of related works in the field of certification automation. 

\cite{stephanow2016testbased,kunz2017process,stephanow2017performance} as well as \cite{anisetti2020} laid the groundwork for automation of certifications by describing the necessary processes and terminology. \cite{anisetti2023} later extended their framework to also include the notion of \textit{continuous}. All the aforementioned approaches heavily rely on a tight coupling between evidences and test cases that generate them. Often a particular test case needs to be generate to collect one specific (type of) evidence.

The EU2020 project MEDINA \cite{medina2021} established a common framework for continuous auditing, but left several challenges unsolved, including the harmonization of evidence collection across the complete service layer. Similarly, \cite{banse2023semanticevidence} formulated the idea of so-called ``semantic evidence'', which can potentially span all the layers of a cloud service. However, in their works they only focus on the infrastructure and not on the complete service including documentation and data. 

\cite{deimling2023amoe} presented research on how evidence of organisational processes can be gathered from documents, but did not harmonize their approach with evidence gathered from technical layers. Other approaches tried to integrate interactive elements, such as chat bots into the approach of continuous monitoring \cite{chatops2022}.

\section{\uppercase{Conclusions and Future work}}
\label{sec:conclusion}

This paper presented the proposed EMERALD approach to support current challenges in the continuous security certification of Cloud Computing services and realize the concept of \textit{certification as a service}. With EMERALD, we expect to significantly decrease the time needed to (re-)certify, select and evaluate new cloud-based services and to facilitate the integration of new services that are not on premise, but offered by different and also smaller providers. 

The EMERALD audit suite for Certification as a Service (CaaS) will contribute to build a strengthened ecosystem of European certification stakeholders enhancing the ability of cloud services providers to maintain regulatory compliance towards the latest security schemes and the upcoming ones and boosting users of cloud services to adopt of European secure solutions, increasing European digital sovereignty in the field of the secure cloud continuum.

The EMERALD project\footnote{\url{https://www.emerald-he.eu}} started in November 2023 and will last 36 months. At the time of writing this paper, the reference architecture for the EMERALD CaaS framework has been designed, and the first versions of the components are available. These initial versions will be integrated into the EMERALD CaaS solution and validated in the four described pilots during 2025.

\section*{\uppercase{Acknowledgements}}

This work has received funding from the European Union’s Horizon Europe Innovation programme under grant no. 101120688.

\bibliographystyle{apalike}
{\small
\bibliography{paper}}



\end{document}